# Analyzing Travel Time Reliability of a Bus Route in a Limited Data Set Scenario: A Case Study

**Ashwini B P*[1], R Sumathi[2], Sudhira H S[3]**

**Abstract**: In this information era commuters prefer to know a reliable travel time to plan ahead of their journey using both public and private modes. In this direction reliability analysis using the location data of the buses is conducted in two folds in the current work; (i) Reliability analysis of a public transit service at route level, and (ii) Travel time reliability analysis of a route utilizing the location data of the buses. The reliability parameters assessed for public transit service are headway, passenger waiting time, travel speed, and travel time as per the Service Level Benchmarks for Urban Transport by the National Urban Transport Policy, Government of India. And travel time reliability parameters such as Buffer Time Index, Travel Time Index, and Planning Time Index are assessed as per Federal Highway Administration, Department of Transportation, U S. The study is conducted in Tumakuru city, India for a significant bus route in a limited data sources scenario. The results suggest that (i) the Level of Service of the public transit service needs improvement. (ii)around 30% excess of average travel time is needed as buffer time. (iii) more than double the amount of free flow travel time must be planned during peak hours and in the worst case. In the future, the analysis conducted for the route can be extended for citywide performance analysis in both folds. Also, the same method can be applied to cities with similar demographics and traffic-related infrastructure.

*Keywords*: Automatic Vehicle Location, Intelligent transportation, Travel time variability, Travel time reliability

## 1. Introduction

Congestion on roads is a major challenge [1] for mobility [2] in urban areas across the globe. The extensive use of private vehicles [3] is one of the major causes of congestion that hampers the overall mobility of a city. Also, excessive use of private vehicles gives rise to, more greenhouse gases and leads to global warming [4]. To avoid these, an alternative mode of transport such as public transit [5] has to be promoted. The most common public transit service available in urban areas/cities in countries like India is buses. Bringing in a modal shift of commuters from private mode to public transit mode is a colossal task, and the role of the transit operations planners [6] is significant in this direction. Transit operations planners are the major stakeholders responsible for the optimization of transit operations like scheduling routes, monitoring service, evaluation, etc. Optimizing transit operations and improving the Level of Service (LoS) [7] will attract more passengers to public transit.

With the implementation of the Intelligent Transportation System (ITS), attempts are being made worldwide to provide a better LoS to commuters. ITS [8] intends to provide services that are innovative such as; information systems [9], navigation systems, adaptive traffic management, incident management, integrated transport management, etc. The growth of Information and Communication Technology (ICT) [10] [11] has opened a variety of data sources [12] related to transportation such as Global Positioning System(GPS), smart cards, automatic fare collection, mobile phone footprints, floating car data, crowdsourced data, etc.

Researchers across the globe have studied various aspects of travel time[13][14] in cities. Some of the most researched issues are regarding reliability [15] in different traffic scenarios [16][17], which are driven by variability[18][19] in various spatial-temporal scales. Reliability of travel time is a major issue and driving force for the modal shift of commuters. Reliability of public transit service is characterized by providing [20] updated information with a consistent travel time [21], minimum waiting time [22] at the bus stop, optimal dwell time [23] [24], adherence with the schedule, optimized operations [25], headways, and regularity [26] between successive service, etc.

Several works assess their reliability features based on the standard defined by Federal Highway Administration (FHWA), United States[27]. Parameters such as buffer time, Buffer Time Index (BTI), Planning Time Index (PTI), and Travel Time Index (TTI) as defined by FHWA are assessed for a selected route in Mysore and other cities of India [28][29]. Also, several works have made effective use of GPS-based data along with other supplementary data sets to assess the reliability of routes with heterogeneous composition[15][30][31][32]. Authors in [25] propose a method for bus route reliability assessment based on the copula function in a case study conducted in China. In [33] the authors used location data of a bus route in Melbourne, Australia, and flow data using loop detectors are used in the framework developed to predict the variance and mean travel time.

[1] *Siddaganga Institute of Technology,Tumakuru, Karnataka, India-572103*
*ORCID ID : 0000-0001-9511-7292*
[2] *Siddaganga Institute of Technology,Tumakuru, Karnataka, India-572103*
*ORCID ID : 0000-0003-0578-3633*
[3] *Gubbi Labs LLP, Gubbi, Karnataka, India-572216*
*ORCID ID : 0000-0002-6568-2327*
*\* Corresponding Author Email: ashvinibp@sit.ac.in*

Authors in [34] assess the impact of external factors on the travel time of public transit buses on the roads of Tri-City Agglomeration in northern Poland and conclude that for developing models to estimate the travel time of public transit vehicles, the sections of the networks have to be analyzed for traffic behavior and the available infrastructure taking into account the dwell time of vehicles. Authors in [35] have conducted an innovative study to assess the quality of service parameters for taxi and ride-hailing and concluded that ride-hailing service is better than taxis in waiting time, cost, and travel time in Jakarta Greater area, Indonesia using survey data. Overall, it is inferred from the existing literature that several research works are conducted using statistical, analytical, and global standard methods, using GPS, smart cards, passenger counters, Lidar and Radar measurements [36], and weather data [37].

*Motivation and objectives:* Most of the existing works are conducted in metropolitan, tier-1, and tier-2 cities with mature traffic-related infrastructure and multiple data sources. In the current scenario, most of the population resides in small cities and urban areas, and effective management of these areas is vital, but the lack of infrastructure is a challenge. Hence with the available location data, a few reliability parameters are assessed in Tumakuru city in two folds as follows

(i) Reliability analysis of a public transit service to assess headway, passenger waiting time, and travel speed and time for the study route
(ii) Travel time reliability analysis of a route considering the buses as probe vehicles to assess BTI, PTI, and TTI.

The findings of this research will emphasize the applications of location data of transit buses in assessing the reliability which was otherwise a tedious field study. Institutions can employ similar methods to assess reliability periodically in a cost-effective manner. The study area, methods followed, and results are discussed in future sections.

## 2. Study Area and Data

The case study is conducted in Tumakuru city, a small-scale city with 370 thousand population. The buses are the only sources of local mass transit in the city. Tumakuru city service is operated by Karnataka Road Transport Corporation (KRTC). Currently, fifteen routes are operating with varying route lengths of 5-15 km. The bus network is connected by segments of national and state highways, and arterial and sub-arterial roads. All city service buses are equipped with a GPS-enabled bus tracking system. The location data of the buses in March 2021 have been used for this analysis. Sample logs are given in Fig. 1.

| vehicleregno | received_date | lattitude | longitude | speed | heading | odometer |
|---|---|---|---|---|---|---|
| KA-06-F-0726 | 08-03-2021 14:43:38 | 13.34278 | 77.09951 | 0 | 265.72 | 19529.82 |
| KA-06-F-0726 | 08-03-2021 14:43:39 | 13.34303 | 77.09927 | 5 | 356.64 | 19529.85 |
| KA-06-F-0726 | 08-03-2021 14:43:49 | 13.34319 | 77.09921 | 5 | 330.15 | 19529.88 |
| KA-06-F-0726 | 08-03-2021 14:43:59 | 13.34326 | 77.09899 | 10 | 275.4 | 19529.9 |
| KA-06-F-0726 | 08-03-2021 14:44:09 | 13.34314 | 77.09875 | 12 | 196.54 | 19529.93 |
| KA-06-F-0726 | 08-03-2021 14:44:19 | 13.34277 | 77.09859 | 17 | 198.17 | 19529.98 |
| KA-06-F-0726 | 08-03-2021 14:44:29 | 13.34234 | 77.09843 | 17 | 198.25 | 19530.04 |
| KA-06-F-0726 | 08-03-2021 14:44:39 | 13.34193 | 77.09829 | 16 | 195.14 | 19530.08 |
| KA-06-F-0726 | 08-03-2021 14:44:49 | 13.34148 | 77.09815 | 20 | 196.63 | 19530.13 |
| KA-06-F-0726 | 08-03-2021 14:44:59 | 13.34101 | 77.09801 | 18 | 196.83 | 19530.18 |
| KA-06-F-0726 | 08-03-2021 14:45:09 | 13.34054 | 77.09789 | 16 | 197.36 | 19530.24 |
| KA-06-F-0726 | 08-03-2021 14:45:19 | 13.34016 | 77.09776 | 17 | 196.39 | 19530.28 |
| KA-06-F-0726 | 08-03-2021 14:45:29 | 13.33975 | 77.09765 | 12 | 194.48 | 19530.33 |
| KA-06-F-0726 | 08-03-2021 14:45:39 | 13.33946 | 77.09758 | 6 | 187.42 | 19530.36 |
| KA-06-F-0726 | 08-03-2021 14:45:49 | 13.33946 | 77.09758 | 0 | 187.42 | 19530.36 |

**Fig. 1** Sample logs of route Tumakuru Bus Stand to Kyathasandra

Route number 201: Tumakuru Bus Stand (TBS) – Kyathasandra (KYA) is selected for the study. The route is further divided into four segments based on the land use pattern. Segment one is Central Business District (CBD), the second segment is Inner City (IC), whereas segment three is the Inner Suburban (ISU) area. The fourth segment is the Outer Suburban (OSU) area, using the national highway. The information on route 201 is given in table 1, and the route map is in Fig 2. The plot of the travel time and the travel speed at the route level is presented in Fig. 3.

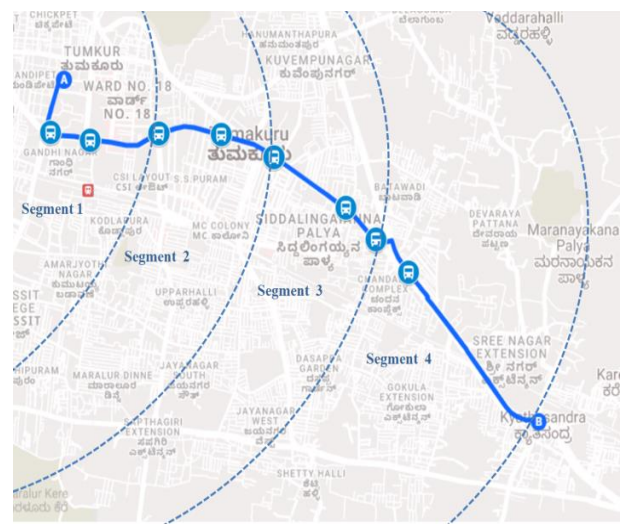

**Fig. 2** Route map of route Tumakuru Bus Stand to Kyathasandra

Table 1: Route information

| Route parameters | Route overall | Segment 1 | Segment 2 | Segment 3 | Segment 4 |
|---|---|---|---|---|---|
| Origin –Destination | TBS-KYA | TBS-Bhadramma Choultry | Bhadramma Choultry –SS Circle | SS Circle-Batawadi | Batawadi-Kyathasandra |
| Length | 6.9 kilometers | 1.76 kilometers | 1 kilometer | 2.09 kilometers | 2.05 kilometers |
| Number of bus stops | 9 | 3 | 2 | 3 | 1 |
| Number of signalized intersections | 6 | 3 | 1 | 1 | 1 |
| Number of lanes | 2 | 2 | 2 | 2 | 3 |
| Land use pattern | - | CBD | IC | ISU | OSU |

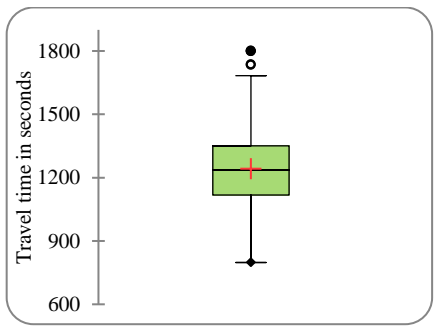

(a)

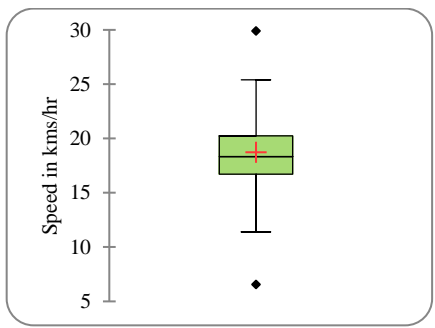

(b)

**Fig. 3** Box plots (a) travel time (b) travel speed of the TBS-KYA route during the study period

## 3. Methods

### 3.1. Public Transit Service Reliability Analysis

Public transit service performance is evaluated based on four components; convenience, comfort, reliability, and security [38]. Reliability is one of the important components that in turn is based on the variability of travel time, passenger waiting time, headway, and punctuality [39]. In this context, with the available location data and the schedule, a few public transit reliability attributes such as headway, passenger waiting time, and travel speed and time for the route under study are assessed. The analysis is conducted based on the Service Level Benchmarks(SLBs) [40] for Urban Transport defined by National Urban Transport Policy (NUTP) - Ministry of Urban Development (MoUD), Government of India.

#### 3.1.1. Headway

It is the time elapsed between two consecutive buses to a common destination at a particular bus stop [41]. The headways at the origin of the route on 10 weekdays are extracted from the location data. Aggregates of the number of buses departing at each Departure Time Window (DTW) between 5:00 in the morning to 21:00 are compared with the scheduled headway. According to the schedule, there are 36 trips for the TBS-KYA route. The comparison is given in Fig. 4 it is observed that there are few DTWs during which the headways differ (violating schedule), illustrating dynamic adjustments in the timetable. The overall headway of the route is estimated using (1).

$$Headway = 1/n * \sum_{i=1}^{n-1}(ST_{i+1} - ST_i) \qquad (1)$$

Where the number of trips along the route is $n$, $ST_i$ is the start time of the trip $i$. The average headway of the route is 25 minutes. For estimating the headway during peak hours, the number of buses scheduled during peak hours is divided by minutes of peak hours. For the current study, 10:00 AM -11:00 AM is considered as peak hour as per [42] [43], three buses are scheduled during peak hour, therefore headway during peak hours is 20 minutes

**3.1.2.** Passenger waiting time: It is the total time, the public transit users spend at the bus stop for the arrival of the bus for the desired route. Passenger waiting time is estimated based on the headway, with the assumption that passengers' arrival rate is uniform at the bus stops. As per the SLBs [40] by NUTP - MoUD, Govt. of India, the Average Passenger Waiting Time (APWT) is estimated using (2).

$$APWT(route) = 1/2(Headway\ of\ route) \quad (2)$$

The APWT of the route TBS-KYA is 12.5 minutes overall, and 10 minutes during peak hours. The APWT of each DTW is shown in **Fig. 5**. According to SLBs, the LoS is at level 3 out of 4 levels. The reference LoS for passenger waiting time according to SLBs of NUTP is given in Table 2.

**Table 2:** Reference table for passenger waiting time[40]

| Level of Service | Average waiting time for public transit users |
|---|---|
| 1 | <=4 mins |
| 2 | 5 - 6 mins |
| 3 | 7 -10 mins |
| 4 | >=11 mins |

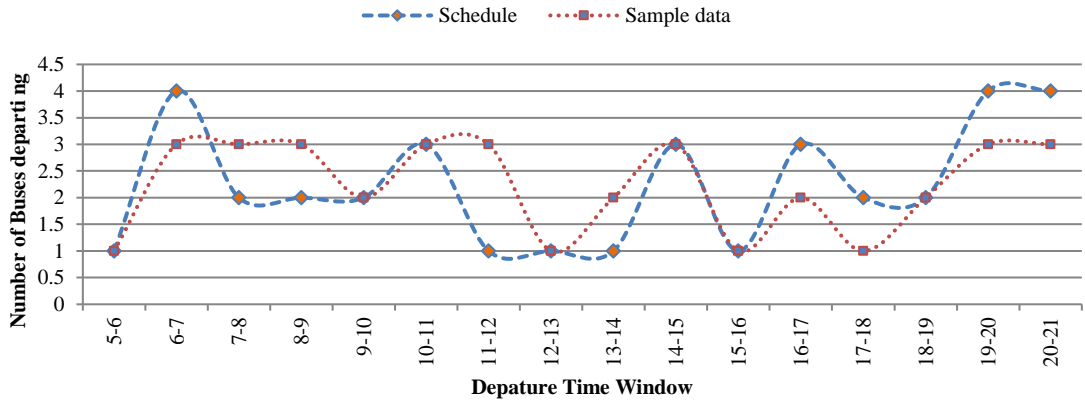

**Fig. 4** Headway of the buses

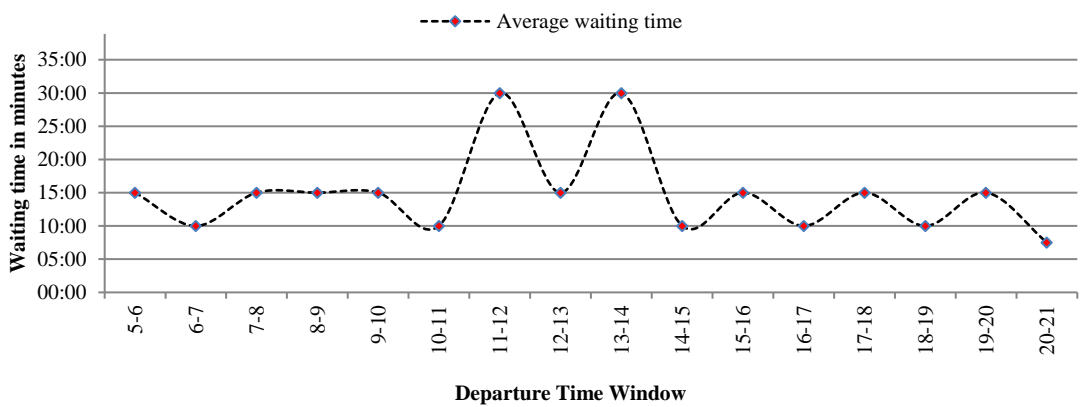

**Fig. 5** Average waiting time at each DTW

### 3.1.3. Travel time

The total travel time from a passenger's point of view is the time from which the passengers start towards the bus stop to the time to reach the final venue. From the service provider's point of view, the travel time is the time taken from the departure of the bus from origin to destination. The analysis of the travel time in the current study is from a service provider's point of view.

*Free-flow travel time vs. Running time:* A free-flow scenario is when there are zero delays i.e., no congestion, waiting times, or disturbances for travel. The trips during 5:00 - 6:00 AM are considered for analyzing the free-flow travel speed and time of the public transit buses. Running time is the travel time of the bus without any delay such as dwell time and intersection delay but in the presence of normal traffic conditions along the route. Trips on weekdays during peak hours excluding the time for acceleration, deceleration, dwell time, and intersection delay [44] are used for analysis. A total of 40 trips, 20 during 5:00 -6:00 AM and 20 during peak hours on weekdays are extracted from location data. The estimations of free flow travel time and speeds, and running time and speeds are summarised in Table 3. An excess running time of around 300 seconds is observed for running time, which contributes to a 52.5% increase in travel time as compared to the free-flow travel time.

*Private mode vs. Public Transit mode:* The most common private vehicle used in Tumakuru city is two-wheelers. According to [43], the mode share of two-wheelers was 60% in 2012. In this context, the travel time of two-wheelers is compared against the bus travel time along the route. The comparison of the private with the public mode can be conducted in many folds [45] such as security, cost, speed, travel time, freedom, comfort, etc. In this study, the comparison is limited to the speed of the vehicle and the travel time along the bus route. 10 two-wheeler trips during peak hours of weekdays were conducted during the study period (March 2021) to analyze the two-wheeler travel time in Tumakuru city. The observations are tabulated in Table 4. An excess travel time of around 370 seconds is observed when public transit buses are used to travel.

SLBs by NUTP, MoUD, India has defined the LoS for transit speed along the major corridors using the motorized personal vehicles and public transit service, the reference is given in Table 5. Based on that, the LoS of personal vehicle speed considering two-wheelers (LoS1) is at level 2, and the LoS of public transit buses (LoS2) is at level 3 (using a weighted average of length). According to column 4 of reference Table 5, LoS1+LoS2 is 5 therefore the overall LoS is at level 3.

**Table 3:** Comparison of free flow and running speeds and travel time

| Segment | Segment Length in kilometers | Free flow Bus speed | Free flow travel time in seconds | Running speed | Running travel time in seconds | Excess running time | Percentage increase in travel time |
|---|---|---|---|---|---|---|---|
| S1 | 1.76 | 38 | 181 | 19 | 333 | 152 | |
| S2 | 1 | 40 | 93 | 28 | 128 | 35 | |
| S3 | 2.09 | 41 | 188 | 27 | 278 | 90 | 52.50% |
| S4 | 2.05 | 49 | 157 | 36 | 205 | 48 | |
| Route | 6.9 | 42 | 619 | 28 | 944 | 325 | |

**Table 4:** Comparison of two-wheeler and bus travel along the route

| Segments | Segment Length in km | Two-Wheeler | | Bus | | Excess travel time by bus |
| | | Mean Speed in km/hour | Mean Travel time in seconds | Mean Speed in km/hour | Mean Travel time in seconds | |
|---|---|---|---|---|---|---|
| S1 | 1.76 | 17 | 375 | 12.5 | 500 | 125 |
| S2 | 1 | 25 | 146 | 13.5 | 280 | 134 |
| S3 | 2.09 | 30 | 250 | 17.5 | 300 | 50 |
| S4 | 2.05 | 40 | 185 | 30 | 250 | 65 |

**Table 5.** Reference table for motorized transit speed [40]

| LoS | Average speed of private vehicles: LoS1 | Average speed of buses: LoS2 | Overall LoS2+LoS3 |
|---|---|---|---|
| 1 | >=30 | >=20 | 2 |
| 2 | 25-29 | 15-19 | 3-4 |
| 3 | 15-24 | 10-14 | 5-6 |
| 4 | <=14 | <=9 | 7-8 |

### 3.2. Travel Time Reliability Analysis of a Route

The FHWA is a part of the Department of Transportation, US and it specializes in transportation. FHWA has defined reliability parameters for a road network. Considering the buses as probe vehicles a few reliability parameters are assessed using only location data in the study area. The assessment of these parameters can be handy for several other study areas that lack other data sources for reliability analysis. In this work parameters namely buffer time index, travel time index, and planning time index are assessed. The 95[th] Percentile Travel Time (95[th] PTT) needed for calculating the indexes is extracted from the location data of trips used in the previous section. The 95th PTT indicates the worst travel time on a heavy travel day. The Free Flow Travel Time (FFTT) is estimated using the speed limit and distance of each section. The average running time of the buses estimated in the previous section is considered the Average Travel Time (ATT) in the current section.

Buffer Time Index: It is the extra time that might be added during travel is called buffer time. This buffer time includes expected delays [46]. BTI is expressed in terms of percentage, indicating the percentage excess buffer time needed to be planned by the commuters to arrive on time at least 95% of the time under normal delays. The formula to compute BTI is given in (3).

$$BTI = ((95th\ PTT - ATT)/ATT) * 100 \quad (3)$$

Planning Time Index: The PTI gives the travel time that must be planned including adequate delays [47] including expected and unexpected delays to arrive at the destination on time. It is a ratio of 95th PTT to that of the FFTT, as given in (4). It is an important parameter that is useful for planning trips where the commuters arrive as per plan in the worst case in 95% of cases.

$$PTI = (95th\ PTT/FFTT) \quad (4)$$

Travel Time Index: It is the ratio of ATT during peak hours to FFTT. i.e., TTI indicates the excess travel time (average) for trips during peak hours as against the FFTT. The formula to estimate the TTI is given in (5).

$$TTI = (ATT/FFTT) \quad (5)$$

The reliability measures are estimated based on equations (3)-(5), and the estimates are summarized in **Table 6**. It is observed from the results in **Table 6** that, segment 2 has the highest BTI (40.63 %) and segment 1 has the highest PTI (3.22), and TTI (2.52) indicating low reliability as compared to other sections. Overall, it is concluded that the travel time of passengers in the study route is high and reliability measures are low.

The reliability parameters are applied for the data of Departure Time Windows (DTW) from 7_8 to 19_20 at the route level. The plot of ATT, 95th PTT, and FFTT is presented in Fig. 6, and PTI and the TTI are presented in Fig. 7.

**Table 6:** Travel time reliability analysis

| Segment | Free Flow Travel Time | 95th Percentile Travel Time | Average Travel Time | Buffer Time Index | Planning Time Index | Travel Time Index |
| --- | --- | --- | --- | --- | --- | --- |
| S1 | 132 | 425 | 333 | 27.63 | 3.22 | 2.52 |
| S2 | 90 | 180 | 128 | 40.63 | 2.00 | 1.42 |
| S3 | 120 | 330 | 278 | 18.71 | 2.75 | 2.32 |
| S4 | 90 | 280 | 205 | 36.59 | 3.11 | 2.28 |
| ROUTE | 432 | 1215 | 944 | 28.71 | 2.81 | 2.19 |

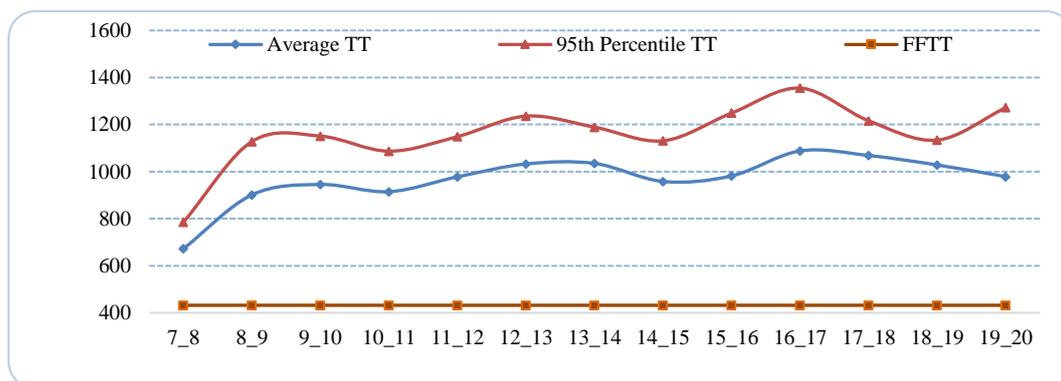

**Fig. 6** ATT, FFTT and 95th PTT at each DTW

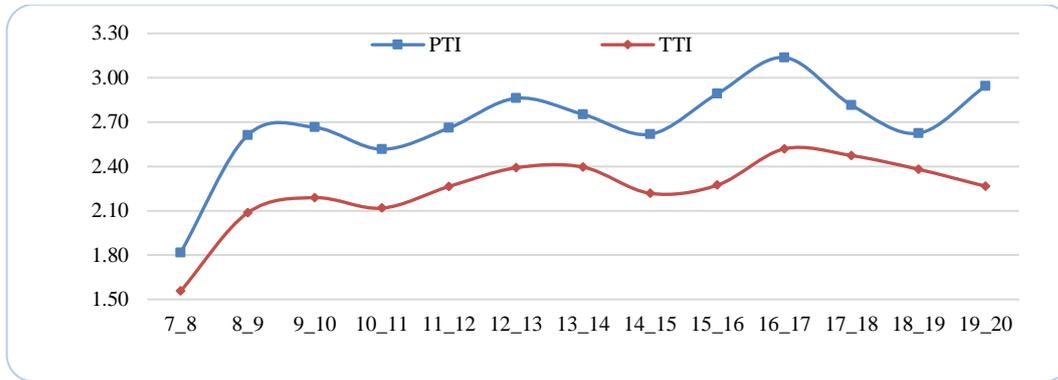

**Fig. 7** The PTI and TTI at each DTW

## 4. Discussion

Travel time reliability [48] is an important feature for commuters and recognizing the variability in travel time and speed can serve to improve it. In heterogeneous traffic conditions with common lanes for general traffic and buses, gaining insights into the travel time behavior of public transit buses is a challenge.

The variability in headway is analyzed, and it is 25 minutes (average) throughout the day and 20 minutes during peak hours. Headway is a major factor that influences reliability, improving the headway can influence the modal shift of the commuters. The average waiting time of the passengers is 12.5 minutes overall and 10 minutes during peak hours. According to the Service Level Benchmarks [40] defined by the National Urban Transport Policy [49], Ministry of Urban Development, Government of India, it is recommended for cities with million-plus populations to maintain a maximum waiting time of 12 minutes. Measures to improve the headway and passenger waiting time are to be considered with priority to sustain and further improve the current scenario as per the NUTP.

A comparison of travel time and speeds of public transit buses with the two-wheeler is conducted and the overall LoS as per SLBs by NUTP is at level 3. Level 3 demonstrates slowness majorly because of high signal density [50], congestion at critical intersections, and inappropriate timing of signals [51] according to NUTP. An average excess travel time of 374 seconds during peak hours is observed as compared to two-wheelers, which account for around 40% excess travel time. This is a major reason for commuters to use the private mode. To overcome this problem, a reliable information system along with deep coverage of services in the network is recommended. Also planning signal priority for public transit buses at intersections is suggested

The free-flow bus speeds and travel time are compared with the running speed and travel times. The results emphasize that there is a total of 52.5% increase in travel time. The bus route is mixed mode and the presence of other vehicles on road, the stochastic behavior of traffic, delays at the intersections, and dwell time reduce the bus speed and increase the travel time. A dedicated lane [20] for buses and prioritizing buses at the intersections [52] [53] can resolve most of the mentioned problems.

Based on the definitions provided by the FHWA, a few reliability parameters namely, BTI, TTI, and PTI are assessed. The results demonstrated an average BTI of 30%, PTI of 2.78, and TTI of 2.14 which are high compared to the results presented in the existing works[15][28] for other cities of India. The range of the reliability scores demonstrates unique travel time behavior in each segment, and to improve the reliability of the route, the treatments have to be provided at the segment level. The route segments belong to the National highway, State highway, Municipality, and Urban development authority. This needs integrated planning and common guidelines. The MoUD, India has suggested the 'Urban Roads Code' [54] for this purpose, and these institutions have to collaborate in this direction to handle the situation.

The analysis of reliability parameters conducted in the study is limited to a major route of the city. This can be extended to assess all other routes of the city and assess the service at the city level. Unlike big cities with multi-mode public transit services such as metro trains, local trains, buses, etc. Tumakuru has only buses as its public transit service mode. There are more than 100 cities like Tumakuru in India, and cities in other Asian countries, the issues, solutions, and recommendations presented to this city can be extended to other cities.

## 5. Conclusion

The reliability of travel is vital for commuters to plan trips. As public transit services are the most used mode of travel in most small cities and urban areas, the reliability of transit services is of equal importance. In this regard, reliability analysis is conducted in two folds; (i) reliability analysis of the public transit service of a study route and (ii) Travel time reliability analysis of a route considering the bus as a probe vehicle. The study is carried out in Tumakuru city, India. Analysis of a few parameters such as headway, passenger waiting time, and speed & travel time in free flow, running, and peak hours scenario is conducted and assessed based on the Service Level Benchmarks suggested by National Urban Transport Policy, MoUD, Govt. of India. The headway is 25 minutes overall and the average waiting time during peak hours is 10 minutes. Excess travel of 40% is observed as compared to a two-wheeler and an excess travel time of 52.5% is observed when compared with a free-flow travel time. According to the benchmarks, the LoS of all the parameters needs improvement.

A few reliability parameters defined by FHWA namely Buffer Time Index (BTI), Travel Time Index (TTI), and Planning Time Index(PTI) of the bus route is estimated. The scores of the reliability parameters of the bus route suggest that around 30% excess time is needed as buffer time as compared to the average travel time. The PTI is 2.78 indicating travelers need to plan for an excess of 2.78 times the Free Flow Travel Time (FFTT) in the worst case, and TTI is 2.14 indicating commuters should plan for 2.14 times the FFTT during peak hours. The values indicate the reliability is low compared to other cities[15] [28]. The analysis was conducted using limited data such as location and schedule data, highlighting the application of the location data in a cost-effective way for locations that lack additional data sources and traffic-related infrastructure. There are several similar cities like Tumakuru in India, and other countries, and the analysis and recommendations presented can be extended to other cities in the future.

## 6. Acknowledgments

The authors are grateful to the authorities of Tumakuru Smart City Limited for providing the essential data (automatic vehicle location logs) of the Tumakuru city transit service buses.

**Author contributions**

**Ashwini B P:** Conceptualization, Methodology, Data collection and pre-processing, Writing-Original draft, **R Suimathi:** Methodology, Data collection, Writing-reviewing, and editing **Sudhira H S:** Conceptualization, Visualization, Investigation, Writing-Reviewing, and Editing.

**Conflicts of interest**

The authors declare no conflicts of interest.